\documentclass[notitlepage,prd,superscriptaddress,nofootinbib]{revtex4}
 \usepackage{color}
\usepackage{amsmath}
\usepackage{amssymb}
\usepackage{amscd}
\usepackage{latexsym}
\usepackage{hyperref}
\def\a{\alpha}
\def\b{\beta}
\def\g{\gamma}

\def\de{\delta}

\def\h{\eta}
\def\th{\theta}

\def\vp{\varphi}

\renewcommand{\and}{{\quad{\rm and}\quad}}

\def\DH{\rm I\kern-1.5pt\rm H\kern-1.5pt\rm I}

\newcommand{\ba}{\begin{array}}
\newcommand{\ea}{\end{array}}
\newcommand{\be}{\begin{equation}}
\newcommand{\ee}{\end{equation}}
\newcommand{\bea}{\begin{eqnarray}}
\newcommand{\eea}{\end{eqnarray}}
\newcommand{\bi}{\begin{itemize}}
\newcommand{\ei}{\end{itemize}}
\def\bv{{\bar v}}
\def\bz{{\bar z}}

\usepackage{mathtools}

\DeclarePairedDelimiter{\diagfences}{(}{)}
\newcommand{\diag}{\operatorname{diag}\diagfences}
\begin{document}
\title{K\"ahler geometry for $su(1,N|M)$-superconformal mechanics}
\author{Erik Khastyan}
\email{khastyanerik@gmail.com}
\affiliation{Yerevan Physics Institute, 2 Alikhanian Brothers St., Yerevan  0036 Armenia}
\author{Sergey Krivonos}
\email{krivonos@theor.jinr.ru}
\affiliation{Bogoliubov Laboratory of Theoretical Physics, Joint Institute for Nuclear Research, Dubna, Russia}
\author{Armen Nersessian}
\email{arnerses@yerphi.am}
\affiliation{Yerevan Physics Institute, 2 Alikhanian Brothers St., Yerevan  0036 Armenia}
\affiliation{Bogoliubov Laboratory of Theoretical Physics, Joint Institute for Nuclear Research, Dubna, Russia}
\affiliation{Institute of Radiophysics and Electronics, Ashtarak-2, 0203, Armenia }
\begin{abstract}
We suggest the $su(1,N|M)$-superconformal mechanics formulated in terms of phase superspace given by the non-compact analogue of complex projective superspace.
We parameterized this phase space   by the specific coordinates allowing to interpret it as a higher-dimensional super-analogue of the Lobachevsky plane parameterized by lower half-plane (Klein model). Then we introduced  the canonical coordinates  corresponding to the known separation of  the "radial" and "angular" parts of (super)conformal mechanics. Relating the "angular" coordinates with action-angle variables we demonstrated that proposed scheme allows to construct the $su(1,N|M)$ supeconformal extensions of wide class of superintegrable systems.  We also proposed  the superintegrable oscillator- and Coulomb- like systems   with a $su(1,N|M)$ dynamical superalgebra, and found that oscillator-like systems admit  deformed $\mathcal{N}=2M$ Poincar\'e supersymmetry, in contrast with Coulomb-like ones.
\end{abstract}

\maketitle
\section{Introduction}\noindent

K\"ahler manifolds are the Hermitian manifolds which possesses the symplectic structure obeying  the specific compatibility condition with the Riemann (and/or complex) structure \cite{arnold}.
Being highly common objects  in almost all areas of theoretical physics, these manifolds   usually appear  as configuration spaces of the particles and fields. Only in  a limited number of physical problems they appear as  phase spaces, mostly for the description of various generalizations of tops,   Hall effect (including its higher-dimensional generalizatons, see, e.g. \cite{nair} and refs therein),    etc.
Respectively, the number of the known  nontrivial (super)integrable systems with K\"ahler phase spaces is very restricted, and their  study  does not attract much attention.
The widely known integrable model  with K\"ahler phase  space  extensively studying   nowadays  is   compactified   Ruijsenaars-Schneider model  with excluded center of mass, whose phase space is   complex projective space \cite{RS}.

On the other hand, there are some indications that K\"ahler phase spaces can  be useful for the study of conventional Hamiltonian systems, i.e. for the systems formulated on cotangent bundle of Riemann manifolds.
 A very simple example of  such system is one-dimensional conformal mechanics formulated  in terms of  Lobachevsky plane (``noncompact complex projective plane")  treated as a phase space \cite{lobach}. Such description, being quite elegant,  allows  immediate  construction of    $\mathcal{N}=2M$ superconformal  extension associated with $su(1,1|M)$ superalgebra.
Recently the similar formulation  of  some  higher-dimensional systems  was given in terms of
 $su(1,N)$-symmetric K\"ahler phase space treated  as the non-compact version of complex projective space \cite{kns}.
In such approach    all symmetries of the generic superintegrable conformal-mechanical systems  acquire interpretation in terms of the powers of
  the  $su(1,N)$ isometry generators. The maximally superintegrable generalizations of the Euclidean  oscillator/Coulomb systems  has also been considered, all the symmetries of these superintegrable systems  were expressed via  $su(1,N)$ isometry generators as well.
  However, the supersymmetrization aspects of that system was not considered there at all.
  In the present paper we construct the $\mathcal{N}$-extended superconformal extensions of the systems considered in \cite{kns}, as it was done in \cite{lobach} for one-dimensional case.
Namely, we consider the systems with $su(1,N|M)$-symmetric   $(N|M)_{\mathbf{C}}$-dimensional K\"ahler phase superspace (in what follow  we denote it by $\widetilde{\mathbb{CP}}^{N|M}$) and relate their symmetries with the isometry generators of the super-K\"ahler structure.
We  construct this superspace  reducing the  $(N+1|M)_{\mathbf{C}}$-dimensional complex pseudo-Euclidean  superspace by the $U(1)$-group action and  then identify the reduced phase superspace with  noncompact analogue of complex projective superspace constructed in \cite{jmp}. We parameterize this superspace    by the
complex bosonic variable $w$, $ {\rm Im}\; w < 0$, by the $N-1$ complex bosonic variables $z^\alpha \in [0, \infty),\;
{\rm arg }\;z \in [0; 2\pi)$, and by $M$ complex fermionic coordinates $\eta^A$.
  Thus, it can be considered as the $N$-dimensional extension of the Klein model of Lobachevsky plane \cite{dnf}. This allows us to   connect the complex coordinate $w$ with the
radial coordinate and momentum of the conformal-mechanical system spanned by $su(1,1)$ subalgebra, and separate the $su(1,1)$   generators interpreting them as Hamiltonian, conformal boosts and dilatation operators.
The rest bosonic generators $z^\alpha $ parameterize the angular part of integrable conformal mechanics with Euclidean configuration spaces\footnote{The convenience of the separation of the radial coordinates from the angular one in the study of conformal mechanics and in their supersymmetrization was demonstrated, e.g., in \cite{angular}}.
 Relating the angular coordinates and momenta with the action-angle
variables, we describe all symmetries of the generic
superintegrable conformal-mechanical systems in terms of the powers of the $su(1,N)$ isometry generators.
An important aspect of proposed approach is  the choice of  canonical coordinates   where all fermionic degrees of freedom appear only in the angular part of the Hamiltonian.

Furthermore,   we construct the super-analogues of the maximally superintegrable generalizations of the Euclidean oscillator/Coulomb systems considered in \cite{kns} as follows: we preserve the form of Hamiltonian expressed via generators of $su(1,1)$ subalgebra but extend the  phase space
$\widetilde{\mathbb{CP}}^{N}$ to phase superspace $ \widetilde{\mathbb{CP}}^{N|M}$. As a result, we find that these superextensions preserve all symmetries of the initial bosonic Hamiltonians and possess maximal set of functionally-independent  fermionic integrals, i.e. they remains superintegrable in the sense of super-Liouville theorem.
We also  find, that   the constructed oscillator-like systems (in contrast with Coulomb-like ones)
possess deformed $\mathcal{N}=2M, d=1$ Poincar\'e supersymmetry (see \cite{ivanovsidorov},
and express all  the symmetries of these superintegrable systems via $su(1,N)$ isometry generators as well.\\

The paper organized as follows.\\
In {\sl Section 2} we present the basic facts on K\"ahler supermanifolds and construct, by the Hamiltonian reduction, the non-compact complex  projective superspace $\widetilde{\mathbb{CP}}^{N|M}$ in the parametrization similar to those of Klein model.
In {\sl Section 3}  we analyze  the symmetry algebra of  $\widetilde{\mathbb{CP}}^{N|M}$ and extract from it the $su(1,N|M)$-superconformal systems.
In {\sl Section 4} we introduce the canonical coordinates which naturally split   radial and angular parts of the Hamiltonian and relate the angular part with the  systems formulating in terms of action-angle variables.
In the {\sl Section 5} we construct superintegrable supergeneralizations of oscillator- and Coulomb-like systems. In {\sl Section 6} we represent the K\"ahler structure of phase superspace  in the  Fubini-Study-like form. We conclude the paper by the  outlook and final remarks in {\sl Section 7}.

\section{ Noncompact complex projective superspace}\noindent
The (even) $(N|M)$-dimensional K\"ahler supermanifold  can be defined as a complex  supermanifold  with symplectic structure  given by the expression
\be
\Omega=\imath
(-1)^{p_I(p_J+1)}g_{I{\bar J}}dZ^I\wedge  d{\bar Z}^J,\quad d\Omega=0,
\ee
with  $Z^I$ denoting $N$ complex bosonic coordinates and   $M$ complex fermionic ones. The
$p_I:=p(Z^I)$   is   Grassmanian  parity of coordinate: it is equal to zero for bosonic coordinate and to one for the fermionic one. Through the paper we will use the following conjugation rule:
 $\overline{Z^I Z^J}=\overline{Z}^I \overline{Z}^J$, $\overline{\overline{Z}^I Z^J}= {Z}^I \overline{Z}^J$,$\overline{\overline{Z}^I \overline{Z}^J}= {Z}^I  {Z}^J$, for both bosonic and fermionic variables.

The ``metrics components"  $g_{I {\bar J}}$ can then be locally represented in the form
 \be
g_{I {\bar J}}=  \frac{\partial^L}{\partial Z^I}
        \frac{\partial^R}{\partial {\bar Z}^J}
                   K  (Z,{\bar Z}),
                   \ee
                   where $\partial^{L(R)}/\partial Z^I$ denotes left(right) derivatives.

The Poisson brackets associated with this  K\"ahler structure looks as follows
\be
\{ f,g\}=\imath\left( \frac{\partial^R f}{\partial \bar Z^I}
               g^{{\bar I}J}
            \frac{\partial^L g}{\partial Z^J}
                   -(-1)^{p_Ip_J}
             \frac{\partial^R f}{\partial Z^I}
             g^{\bar JI}
         \frac{\partial^L g }{\partial \bar Z^J}
               \right)
,\quad {\rm  where}\quad
g^{{\bar I}J}g_{J{\bar K}}=\delta^{\bar I}_{\bar K},
\quad  \overline{g^{\bar I  J}}
 = (-1)^{p_Ip_J}g^{\bar J I}, \ee
As in the pure bosonic case, the isometries of K\"ahler manifolds are given by the {\sl holomorphic Hamiltonian vector fields},
\be
\mathbf{V_\mu}:=\{h_\mu(Z, \bar Z),\quad\}= V^I(Z)\frac{\partial^L}{\partial Z^I}+ {\bar V}^I({\bar Z})\frac{\partial^L}{\partial {\bar Z}^I},
\ee
  where $h_\mu(Z, \bar Z)$ are real functions called Killing potentials (see, e.g. \cite{lnp,jmp} for the details).

Our goal is to  study the systems on the K\"ahler phase space  with $su(1,N|M)$ isometry superalgebra. For the construction of such phase space  it is convenient, at first, to present the
 linear realization of $u(1,N|M)$ superconformal algebra  on  the complex pseudo-Euclidean superspace $\mathbb{C}^{1,N|M}$ equipped with the
canonical K\"ahler structure (and thus, by the
canonical supersymplectic structure)and then reduced it by the action of $U(1)$ generator.

It is instructive to present this reduction in details.
Let us equip, at first, the   $(N+1|M)$-dimensional complex superspace with the canonical symplectic structure
\be
\Omega_0=\imath \sum_{a,b =0}^{N}\gamma_{a\bar b}dv^a \wedge d{\bar v}^b +\sum_{A=1}^M d\eta^A\wedge d{\bar \eta}^A,
\ee
 with $v^a, \bar{v}^a$ being bosonic variables, and $\eta^A, {\bar \eta}^A$  being fermionic ones, and with the matrix $\gamma_{a\bar b}$ chosen in the form
\be
\gamma_{a\bar b}=
\left(
\begin{array}{c|c}
\begin{matrix}
0 & -i \\
i & 0
\end{matrix}
&  \\
\hline
 &
\begin{matrix}
-1 &  \\
 & \ddots & \\
 & & -1
\end{matrix}
\end{array}
\right),
\qquad a,b=N,0,1,...,N-1 .
\label{mat0}\ee
With this supersymplectic structure we can associate the  Poisson brackets
 given by the  relations
 \be
 \{v^a,{\bar v}^b\}=-\imath\gamma^{\bar b a}, \quad \{\eta^A, \bar \eta^B\}=\{\bar \eta^B, \eta^A\}=\delta^{A\bar B}, \qquad \gamma^{\bar a b}\gamma_{b \bar c}=\delta^{\bar a}_{\bar c }.
 \label{pb0}\ee
 Equivalently,
\be
\{v^0,\bar v^N \}=1, \qquad \{v^N,\bar v^0 \}=-1, \qquad \{v^\alpha, \bar v^\beta\}=\imath \delta^{\a \bar \b},\qquad
\{\eta^A, \bar \eta^B\}=\{\bar \eta^B, \eta^A\}=\delta^{A\bar B}
\ee
Here we  introduced the indices $\alpha,\beta=1,\ldots, N-1$.

On this superspace we can define the linear Hamiltonian action of $u(1,N|M)=u(1)\times su(1,N|M)$ superalgebra
\begin{align}
&\{h_{a\bar b},h_{c\bar d}\}=-\imath\left(h_{a \bar d}\gamma^{\bar c b}- h_{c \bar b}\gamma^{\bar a d}\right),\quad
\{\Theta_{A\bar a}, {\bar\Theta}_{\bar B b}\}=h_{b \bar a}\delta^{B \bar A}-R_{A\bar B}\gamma^{\bar b a},
\quad \{\Theta_{A\bar a}, h_{b\bar c}\}=-\imath\Theta_{A\bar c}\gamma^{\bar b a},
&\label{ulnm1} \\
&
\{R_{A\bar B},R_{C\bar D}\}=\imath \left(R_{A \bar D}\delta^{B \bar C}-R_{C\bar B}\delta^{D\bar A}\right),\quad \{\Theta_{A\bar a}, R_{C\bar D}\}=-\imath\Theta_{C\bar a}\delta^{D\bar A}, &
\label{u1nm2}\end{align}
where
\be
h_{a\bar b}=\bv^a v^b,\qquad \Theta_{A\bar a}=\bar\eta^A v^a,\qquad R_{A\bar B}=\imath\bar\eta^A\eta^B .
\ee
The $u(1)$ generator defining the center of $u(N.1|M)$ is given by the expression
\be
J=\gamma_{a\bar b}v^a {\bar v}^b+\imath\eta^A\bar\eta^A\;:\{J, h_{a\bar b}\}=\{J, \Theta_{A\bar a}\}=\{J, R_{A\bar B}\}=0.
\label{J}\ee
Hence, reducing the system by the action of this generator we will get the "non-compact" projective super-space $\widetilde{\mathbb{CP}}^{N|M}$ (i.e. the  supergeneralization of noncompact projective space $\widetilde{\mathbb{CP}}^{N}$), which is $(2N|2M)$-(real)dimensional space.

For performing the reduction by the action of  generator \eqref{J} we have to choose, at first,  the  $2N$ real ($N$ complex) bosonic and  $2M$ real ($M$ complex) fermionic functions commuting with $J$.
Then, we have to calculate their Poisson brackets and restrict the latters to the level surface
\be
J=g.
\label{ls}\ee
As a result we will get the Poisson brackets on the reduced $(2N|2M)$-(real) dimensional space, with that $U(1)$-invariant functions playing the role of the latter's coordinates.

The required functions could be  easily found as
\be
w=\frac{v^N}{v^0}, \quad z^\a =\frac{v^\a}{v^0}, \quad \th^A  =\frac{\h^A}{v^0}\;: \qquad \{w, J\}=\{z^a,J\}=\{\th^A, J\}=0,  \qquad {\rm and } \quad c.c. .
\label{rf}\ee
Calculating their Poisson brackets and having in mind the  expression
following from  \eqref{ls},
\be
A:=\left.\frac{1}{v^0\bar v^0}\right\rvert_{J=g}= \frac1g\left(\imath(w-\bar w)-\sum_{\g=1}^{N-1} z^\g\bar z^\g +
\imath \sum_{C=1}^{M} \th^C \bar \th^C \right),
\label{A}\ee
we get   the  reduced Poisson brackets defined by the following non-zero relations (and their complex conjugates)

\be
\{w, \bar w \}=-A(w-\bar w), \quad \{z^\a, \bar z^\b \}=\imath A \delta^{\a \bar \b},
\quad \{\th^A, \bar \th^B \}=A\delta^{A\bar B},\quad
\{w,\bar z^\alpha \}=A\bar z^\alpha, \quad \{w,\bar \th^A \}=A\bar \th^A.
\label{pbb}\ee
These Poisson brackets are associated with the supersymplectic structure
\begin{align}
\Omega =\frac{\imath}{g} &\left[\frac{1}{A^2}dw \wedge d\bar w -\frac{\imath z^\a}{A^2}dw \wedge d\bar z^\a-
\frac{\th^A}{A^2}dw\wedge d\bar\th^A \right.\nonumber \\
&+\frac{\imath \bar z^\a}{A^2}dz^\a\wedge d\bar w + \left(\frac{g\de_{\a\bar \b}}{A}+\frac{\bar z^\a z^\b}{A^2}\right)
dz^\a \wedge d\bar z^\b-
\frac{\imath \bar z^\a \th^A}{A^2}dz^\a \wedge d\bar \th^A \nonumber \\
&-\frac{\bar \th^A}{A^2}d\th^A\wedge d\bar w +\frac{\imath \bar \th^A z^\a }{A^2} d\th^A\wedge d\bar z^\a-
\left.
\left(\frac{\imath g \de_{A\bar B}}{A} +\frac{\bar \th^A \th^B}{A^2}\right)d\th^A\wedge d\bar \th^B
\right].
\end{align}
It is defined by the K\"ahler potential
\be
{{\mathcal{K}}}=-g\log (\imath({{w}}-{\bar{w}})- z^\alpha\bz^\alpha
+\imath\theta^A\bar\theta^A). \label{skah}\ee
In what follows we will  call this space `` noncompact projective superspace $\widetilde{\mathbb{CP}}^{N|M}$ ".
The isometry algebra of this space is $su(1,N|M)$, which can be easily obtained by the restriction of the generators \eqref{ulnm1},\eqref{u1nm2} to the level surface \eqref{ls}.
It is defined  by the following Killing potentials
\begin{align}
&H:=v^N\bar v^N\vert_{J=g}=\frac{w \bar w}{A},
&\quad  &K:=v^0\bar v^0\vert_{J=g} = \frac{1}{A},
&\quad  &D:=(v^N \bar v^0 + v^0 \bar v^N)\vert_{J=g}=\frac{w+\bar w}{A},\label{b1}\\
&H_{\a}:=\bar v^\a v^N\vert_{J=g}=\frac{\bar z^\a w}{A},
&\quad &K_\a:=\bar v^\a v^0\vert_{J=g}=\frac{\bar z^\a}{A},
&\quad &h_{\a \bar \b}:=\bar v^\a v^\b\vert_{J=g} = \frac{\bar z^\a z^\b}{A},\label{b2}\\
&Q_A:=\bar \h^A v^N\vert_{J=g}= \frac{\bar \th^A w}{A},
&\quad &S_A:= \bar \h^A v^0\vert_{J=g}= \frac{\bar \th^A}{A},
&\quad &\Theta_{A \bar \a}:=\bar \h^A v^\a\vert_{J=g}=\frac{\bar \th^A z^\a}{A},\label{b3}\\
&R_{A\bar B}:=\imath\bar \h^A \h^B \vert_{J=g} =\imath \frac{\bar \th^A \th^B}{A}\label{b4}.
\end{align}
Constructed super-K\"ahler structure can be viewed as a higher dimensional
analogue of the Klein model of Lobachevsky space, where the latter is parameterized by the lower half-plane.
One can choose, instead of non-diagonal matrix \eqref{mat0}, the diagonal one, $\gamma_{a\bar b}=diag(1,-1,\ldots, -1)$. In that case the reduced K\"ahler structure will have the Fubini-Study-like form (see Section VI).  In the next Section we will analyze  the   isometry algebra defined by these generators in details.
Presented choice \eqref{mat0} is  motivated by  its convenience for the analyzing superconformal mechanics.
Indeed, in that case the generators \eqref{b1} define conformal subalgebra $su(1,1)$ and are separated from the rest $su(N,1)$ generators. Thus they  can be interpreted as  the Hamiltonian of conformal mechanics, the generator of conformal boosts and the generator of dilatation.

In the next section we will analyze in details these superconformal mechanics and their dynamical  defined by the generators   \eqref{b1},\eqref{b2},\eqref{b3},\eqref{b4}.
  \section{$su(1,N|M)$ superconformal algebra}
The generators (Killing potentials) \eqref{b1},\eqref{b2},\eqref{b3},\eqref{b4} form  $su(1,N|M)$ superalgebra  given by \eqref{ulnm1},\eqref{u1nm2} with $\gamma_{a\bar b}$ defined in \eqref{mat0}. Its explicit expression with separated $su(1,1)$ subalgebra  is represented below.
For the convenience it is  divided  into three sectors: "bosonic", "fermionic" and "mixed" ones.

\subsubsection*{Bosonic sector: $su(1,N)\times u(M)$ algebra }
The bosonic sector is the direct product of the $su(1,N)$  algebra defined by the generators \eqref{b1},\eqref{b2},
and the $u(M)$ algebra defined
 by the R-symmetry generators  \eqref{b4}.
 Explicitly, the $su(1,N)$ algebra is given by the relations
\begin{align}
&\{ H , K \}=-D, \quad \{H , D \}=-2H, \quad \{K , D \}=2K,&\label{bg1}\\
&
\{H , K_\a \}=-H_{\a }, \quad \{H , H_{\a }\}=\{H , h_{\a \bar \b}\}=0,
&\\
&
\{K , H_{\a }\}=K_\a, \quad\{K , K_\a \}=\{K , h_{\a \bar \b}\}=0,
&\\
&
\{D, {K}_\a \}=-K_\a, \quad \{D , H_{\a}\}=H_{\a }, \quad \{D , h_{\a \bar \b} \}= 0,
&\\
&
\{K_\a ,K_\b \}= \{H_{\a},H_{\b }\}=\{K_{\alpha},H_{\beta}\}=0,
&\\
&
\{K_\a , {\overline K}_{\b}\}= -\imath K\delta_{\a \bar \b},\quad
\{H_{\a }, {\overline H}_{\b} \} = -\imath H \delta_{ \a \bar\b}, \quad
\{h_{\a \bar \b}, h_{\g \bar \delta}\}=\imath(h_{\a \bar \delta} \delta_{\g \bar\b}-h_{\g \bar \b}\delta_{ \a \bar\delta}),
&\\
&
\{K_\a , h_{\b \bar \g} \}=-\imath K_\b \delta_{\a \bar \g}, \quad
\{H_{\a},h_{\b \bar \g} \} = -\imath H_{\b} \delta_{ \a \bar\g},\qquad \{K_\a , {\overline H}_\b\}= h_{\a \bar \b}
+\frac{1}{2}\left(I-\imath D\right)\delta_{ \a \bar\b},
&
\end{align}
where
\be
I:=g+\sum_{\gamma=1}^{N-1} h_{\gamma \bar\gamma}+\sum_{C=1}^M R_{C \bar C}
\label{casimir}\ee
The R-symmetry generators form $u(M)$ algebra  and commutes with all generators of $su(1,N)$:
\be
\{R_{A\bar B}, R_{C \bar D}\}= \imath(R_{A\bar D}\delta_{C\bar B}-R_{C\bar B}\delta_{A\bar D})
, \quad\{R_{A\bar B},(H;K;D; K_{\a};H_{\a};h_{\a \bar \b})\}=0.
\ee
It is clear that  the generators $H,D, K$ form conformal algebra $su(1,1)$,
the generators $h_{\alpha\bar\beta}$ form the algebra $u(N-1)$, and all together - the $su(1,1)\times u(N-1)$ algebra.
Notice,  that  the generator  $I$ in \eqref{casimir} defines the Casimir of conformal algebra $su(1,1)$:
 \be
 \mathcal{I}:=\frac{1}{2}I^2=\frac{1}{2}D^2-2HK
 .\label{cas}\ee
Hence, choosing  $H$  as a  Hamiltonian, we get that $H_\alpha$, $h_{\alpha\bar\beta}, R_{A\bar B}$ define its constant of motion.
Similarly, choosing the generator $K$ as a  Hamiltonian, we get that it has constants of motion  $K_\alpha, h_{\alpha\bar\beta}, R_{A\bar B}$.

\subsubsection*{"Fermionic" sector}
The Poisson brackets between fermionic generators \eqref{b3} have the form
\begin{align}
&
\{S_{A},{\overline S}_{ B}\}=K\delta_{A\bar B}, \quad \{Q_{A}, {\overline Q}_{B}\}=H\delta_{A\bar B}, \quad
\{S_{A},{\overline Q}_{ B}\} = -\imath R_{A \bar B} +\frac{\imath}{2}\left(I-\imath D\right)\delta_{A \bar B} ,
&\label{f1}\\
&
\{\Theta_{A \bar \a}, {\overline\Theta}_{B\bar\b}\}= R_{A\bar B}\delta_{\b \bar \a}+h_{\b \bar \a}\delta_{A\bar B},
\quad
\{S_{A},{\overline\Theta}_{B\bar \a }\}= K_\a \delta_{A\bar B}, \quad \{Q_{A}, {\overline\Theta}_{B\bar\a}\}= H_{\a}\delta_{A\bar B},
&\label{f2}\\
&
\{S_{A}, S_B \}=\{Q_{A }, Q_{B}\}= \{\Theta_{A \bar \a}, \Theta_{B \bar \b}\}= \{S_{A},Q_{B}\}=\{S_{A}, \Theta_{B \bar \a}\}=
\{Q_{A}, \Theta_{B \bar \a}\}=0.
&\label{f3}
\end{align}
Hence, the functions $Q_A$ play the role of supercharges  for the Hamiltonian $H$, and the functions $S_A$ define the supercharges  of the Hamiltonian given by the generator of conformal boosts $K$.
\subsubsection*{"Mixed" sector}
The mixed sector is given by the relations
\begin{align}
&
\{H,Q_{A}\}=\{H, \Theta_{A\bar \a}\}=0,\quad \{H,S_{A}\}=-Q_{A },
&\\
&
\{K,S_{A}\}=\{K,\Theta_{A\bar \a}\}=0, \quad\{K,Q_{A}\}=S_{A},
&\\
&
\{D,S_{A}\}=-S_{A}, \quad \{D,Q_{A }\}=Q_{A}, \quad \{D,\Theta_{A\bar \a}\}=0
&\\
&
\{Q_{A}, {\overline K}_\a \}=-\Theta_{A\bar \a}, \quad \{Q_{A}, H_{\a}\}=\{Q_{A}, {\bar H}_{\a}\}=\{Q_{A}, {\bar K}_{\a}\}=\{Q_{A},h_{\a \bar \b}\}=0,
&\\
&
\{S_{A}, {\overline H}_{\a} \}=\Theta_{A\bar \a}, \quad \{S_{A}, {K}_{\a}\}=\{S_{A}, {\bar K}_{\a}\}=\{S_{A}, {H}_{\a}\}=\{S_{A},h_{\a \bar \b}\}=0,
&\\
&
\{\Theta_{A\bar \a}, K_{\b} \}=\imath S_{A} \delta_{\b \bar \a}, \quad
\{\Theta_{A\bar \a},H_{\b}\}=\imath Q_{A}\delta_{\b\bar \a},
\quad \{\Theta_{A\bar\a}, {\bar H}_\alpha\}=\{\Theta_{A\bar\a}, {\bar K}_\alpha\}=0,\quad \{\Theta_{A\bar \a},h_{\b\bar \g}\}=\imath \Theta_{A \bar \g}\delta_{\b \bar \a},\\
&
\{S_{A}, R_{B\bar C}\}=-\imath S_B\delta_{A\bar C}, \quad \{Q_{A}, R_{B\bar C}\}=-\imath Q_{B}\delta_{A\bar C},\quad
\{\Theta_{A\bar \a}, R_{B\bar C}\}=-\imath \Theta_{B\bar \a}\delta_{A\bar C}.\label{mlast}
&
\end{align}
Looking to the all Poisson bracket relations together we conclude that
\begin{itemize}
\item
The  bosonic functions $H_{\alpha}$, $h_{\alpha\bar\beta}$ ,   and the fermionic functions
$Q_A$, $\Theta_{A\bar\alpha}$  commute with the Hamiltonian   $H$  and thus,  provide it by the superintegrability property \footnote{In accord with super-analogue of Liouville theorem \cite{shander}
the system on $(2N.M)$ phase superspace is integrable iff it possess $N$ commuting bosonic integrals (with nonvanishing and functionally independent bosonic parts)  and $M$ fermionic ones };

\item
The  bosonic functions $K_{\alpha}$, $h_{\alpha\bar\beta}$ and the fermionic functions
$S_A$,$\Theta_{A\bar\alpha} $  commute with the generator  $K$. Hence, the Hamiltonian $K$  defines the superintegrable system as well.

\item The triples  $(H, H_{\alpha}, Q_A, )$ and  $(K, K_{\alpha},  S_A, )$ transform  into each other  under the
 discrete transformation
\be
(w, z^\alpha, \theta^A)\to (-\frac1w,\frac{z^\alpha}{w},\frac{\theta^A}{w} )\quad\Rightarrow D\to -D,
\quad\left\{\begin{array}{ccc}(H, H_{\alpha},  Q_A,)&\to &(K, -K_{\alpha}, - S_A),\\
(K, K_{\alpha}, S_A)&\to &(H,H_{\alpha},  Q_A,)
\end{array}\right. .
\label{duality}\ee
\item The functions $h_{\alpha\bar\beta}, \Theta_{A\bar\alpha}$ are invariant under discrete transformation \eqref{duality}. Moreover, they appear to be constants of motion both for $H$ and $K$. Hence, they remain to be constants of motion for any Hamiltonian being the functions of $H,K$.
    In particular, adding  to the  Hamiltonian $H$  the appropriate function of $K$, we get the superintegrable  oscillator- and Coulomb-like systems with dynamical superconformal symmetry (see Section V) .

    \item
  The superalgebra $su(1,N|M)$ admits 5-graded decomposition \cite{5gr,grading}
\be\label{5gr0}
 su(1,N|M)= \mathfrak{f}_{-2} \oplus  \mathfrak{f}_{-1} \oplus \mathfrak{f}_{0} \oplus  \mathfrak{f}_{+1}\oplus  \mathfrak{f}_{+2}
\qquad\textrm{with}\qquad
\left[  \mathfrak{f}_i,  \mathfrak{f}_j\right] \subseteq  \mathfrak{f}_{i+j}
\quad\textrm{for} \ i,j\in\left\{ -2,-1,0,1,2\right\},
\ee
 where $\mathfrak{f}_i=0$ for $|i|>2$ is understood.
The subset
$ \mathfrak{f}_{0} $ includes the   generators  $D, h_{\alpha\bar\beta}, \Theta_{A\bar\alpha},{\overline\Theta}{A\bar\alpha}, R_{A\bar B}$,
the subsets  $ \mathfrak{f}_{-2} $ and   $ \mathfrak{f}_{2} $ contain only
generators $H$  and $K$, respectively,  while the subsets $ \mathfrak{f}_{-1} $ and   $ \mathfrak{f}_{1} $ contain the generators
$H_\alpha, {\bar H}_\alpha, Q_A, {\bar Q}_A$ and $K_\alpha, {\bar K}_\alpha, S_A, {\bar S}_A$.

\end{itemize}

Let us conclude this section by the following remark.
It is easy to see, that the generator \eqref{casimir} commutes the generators $H,D,K, S_A,Q_A, R_{A\bar B}$. Hence,
these generators form superconformal  algebra $su(1, 1|M)$ with central charge $\sqrt{2\mathcal{I}}$ \eqref{cas} (being the Casimir of $su(1,1|M))$ as well)
\begin{align}
&\{ H , K \}=-D, \quad \{H , D \}=-2H, \quad \{K , D \}=2K,\quad \{S_{A},{\overline S}_{ B}\}=K\delta_{A\bar B}, \quad \{Q_{A}, {\overline Q}_{B}\}=H\delta_{A\bar B},\nonumber &\\&
\{S_{A},{\overline Q}_{ B}\} = -\imath R_{A \bar B} +\frac{\imath}{2}\left(\sqrt{2\mathcal{I}}-\imath D\right)\delta_{A \bar B},\nonumber &\\&
\{H,S_{A}\}=-Q_{A },\quad \{K,Q_{A}\}=S_{A},\quad\{H,Q_{A}\}=\{K,S_{A}\}=0,
\quad \{D,S_{A}\}=-S_{A}, \quad \{D,Q_{A }\}=Q_{A},\nonumber &\\&
\{R_{A\bar B}, R_{C \bar D}\}= \imath(R_{A\bar D}\delta_{C\bar B}-R_{C\bar B}\delta_{A\bar D}),\quad \{S_{A}, R_{B\bar C}\}=-\imath S_B\delta_{A\bar C}, \quad \{Q_{A}, R_{B\bar C}\}=-\imath Q_{B}\delta_{A\bar C}.
\label{su11N}\end{align}

In the next section we will express presented $su(1, N|M)$ generators in appropriate canonical coordinates and in this way we will  relate presented formulae with the
superextensions of conventional conformal mechanics.

%
%
%
%

\section{Canonical coordinates and action-angle variables}\noindent

To define  the canonical coordinates we pass from the complex bosonic coordinates $w$, $z^\alpha$
\be
 w=x+\imath y , \quad z^\alpha = q_\alpha {\rm e}^{\imath\varphi_\alpha}, \quad{\rm where}\qquad y <0,\quad  q_\alpha\geq 0,
 \quad \varphi_\alpha\in [0, 2\pi ),\quad q^2:=\sum_{\alpha=1}^{N-1} q_\alpha^2 < -2y.
\ee
Then we  re-define fermionic ones such that the new variables will have canonical Poisson brackets.

For this purpose
  we write down the symplectic/K\"ahler one-form   and identify it with the canonical one
\be
\mathcal{A}= -\frac{g}{2}\frac{dw+d\bar w-\imath (z^\a d \bar z^\a-\bar z^\a dz^\a) +\th^A d\bar \th^A +\bar \th^A d \th^A}{\imath(w-\bar w)-z^\g \bar z^\g + \imath \th^C \bar \th^C}
:= p_x dx+\pi_\a d\vp_\a + \frac{1}{2}\chi^A d\bar \chi^A + \frac{1}{2}\bar \chi^A d\chi^A
\ee
 After some calculations and canonical transformation $(p_x,x) \rightarrow (-\frac{r^2}{2}, \frac{p_r}{r})$, one can obtain
\be
w=\frac{p_r}{r}-\imath \frac{I}{r^2}, \quad
z^\a =\frac{\sqrt{2\pi_\a}}{r}{\rm e}^{\imath \vp_\a}, \quad
\th^A = \frac{\sqrt{2}}{r}\chi^A,
\ee
where $r,p_r, \pi_\alpha, \varphi_\alpha, \chi^A,\bar\chi^A$ are canonical coordinates.
\be
\{r,p_r\}=1,\quad \{\varphi_\b, \pi_\a\}=\delta_{\a \b},\quad \{\chi^A,\bar\chi^B\}=\delta^{A\bar B},\qquad \pi_a\geq 0, \quad \varphi^a\in[0,2\pi ), \quad r>0.
\ee
 They expresses via initial ones as follows
\be
p_r=\frac{w+\bar w}{2}\sqrt{\frac{2}{A}},\quad
r=\sqrt{\frac{2}{A}},\quad
\pi_\a=\frac{z^\a \bar z^\a }{A},
\quad
\varphi_\a= \arg(z^\a),\quad
\chi^A=\frac{\theta^A}{\sqrt{A}}, \quad c.c.,
\label{cc}
\ee
where
\be
I= g+\sum_{\a=1}^{N-1} \pi_\a+ \sum_{A=1}^{M}\imath\bar\chi^A\chi^A\;,\qquad  A:= \frac{\imath(w-\bar w)-z^\g\bar z^\g +\imath \th^C \bar \th^C}{g} = \frac{2}{r^2}.
\ee

In these canonical coordinates the isometry generators read
\begin{align}
&H=\frac{p_r^2}{2}+\frac{I^2}{2r^2}, \quad
 K=\frac{r^2}{2},\quad
 D= p_r r,&\label{can1}
\\
&H_{\a}=\sqrt{\frac{\pi_\a}{2}}{\rm e}^{-\imath \vp_\a}\left({p_r} - \imath \frac{I}{r}\right), \quad
 K_\a =r\sqrt{\frac{\pi_\a}{2}}{\rm e}^{-\imath \vp_\a},\quad
 h_{\a \bar \b}=\sqrt{\pi_\a \pi_\b}{\rm e}^{-\imath(\vp_\a-\vp_\b)},&\label{can2}
\\
&Q_{A}=\frac{\bar \chi^A}{\sqrt{2}} \left(p_r-\imath\frac{\sqrt{2\mathcal{I}}}{r}\right), \quad
S_{A}= \frac{\bar \chi^A}{\sqrt{2}}r,\quad
\Theta_{A \bar \a}=\bar \chi^A \sqrt{\pi_\a}{\rm e}^{\imath \vp_\a},
\quad
R_{A\bar B}=\imath\bar \chi^A \chi^B.&\label{can3}
\end{align}
Interpreting $r$ as a radial coordinate, and $p_r$ as radial momentum, we
  get the superconformal mechanics with angular Hamiltonian given by

\be
 \mathcal{I}=\frac{I^2}{2}:=\frac12\left(I_0+(\bar\chi\chi)\right)^2, \qquad{\rm with}\qquad  I_0:=g+\sum_{\a=1}^{N-1} \pi_\a,\quad (\bar\chi\chi):=\sum_{A=1}^{M}\imath\bar\chi^A\chi^A\;.
\label{casimir2}\ee
 So,  the fermionic part of superconformal Hamiltonian is encoded in its angular part.  \\

 The explicit dependence of the Hamiltonian $H$  and  the  supercharges $Q_A$  on the fermions is as follows
 \be
H=H_0+\frac{I_0(\bar\chi\chi)}{r^2}+\frac{(\bar\chi\chi)^2}{2r^2}, \qquad Q_A=-\frac{\bar \chi^A}{\sqrt{2}} \left(p_r-\imath\frac{I_0}{r}-\imath\frac{(\bar\chi\chi)}{r}\right),\qquad
\ee
while the  dependence of bosonic integrals $H_\a$ on fermions is given by the expression
\be
 H_{\a}= H^0_\alpha - \frac{K_\alpha (\bar\chi\chi)}{2K},
 \ee
 where
 \be
 H_0:=\frac{p^2_r}{2}+\frac{I^2_0}{2r^2},\quad  H^0_\a=\sqrt{\frac{\pi_\a}{2}}{\rm e}^{-\imath \vp_\a}\left({p_r} - \imath \frac{I_0}{r}\right)\; :\quad\{H^0_\a, H^0\}=0.
\ee
So, proposed  superconformal Hamiltonian $H$ inherits   all  symmetries of initial Hamiltonian $H_0$ (given by $H^0_\alpha, h_{\alpha\bar\beta}$).\\

Looking at the functional dependence of the angular Hamiltonian $\mathcal{I}$ from the angular variables $\varphi^\a, \pi_\a$ one can expect that
the set of conformal mechanics admitting proposed $su(1, N|M)$ superconformal extensions seems to be  very restricted. However, it is not the  case, since we it is not necessary to interpret $\varphi^\a$ as a coordinate of the configuration space, and $\pi_\a$ as its canonically conjugated momentum.
Instead, since $\pi_\a$ define a constant of motion of the  bosonic Hamiltonian $H_0$ (and of the respective angular Hamiltonian $\mathcal{I}_0=H_0K/2 -D^2$), we can interpret it as the action variable $I_\a$,  and consider $\varphi^\a$ as  a respective  angle  variable $\Phi_\a$ .

Furthermore,  suppose that
 $\pi_\alpha,\varphi_\alpha$ are related with the  action-angle variables $(I_\alpha,\Phi_\alpha)$ of some $(N-1)$-dimensional angular mechanics by the relations
\be
\pi_\alpha=n_\alpha I_\alpha,\quad \varphi_\alpha=\frac{\Phi_\alpha}{n_\alpha},\qquad {\rm where}\quad n_\alpha\in \mathbb{N}, \qquad \{\Phi_\alpha, I_\beta\}=\delta_{\alpha\beta},\qquad \Phi_\alpha \in[0,2\pi).
\label{paa}\ee
Upon this identification the bosonic part of the angular Hamiltonian \eqref{casimir2} takes a form
 \be
 \widetilde{\mathcal{I}}_0=\frac12\left(g+\sum_{\alpha=1}^{N-1}n_\alpha I_\alpha 
 \right)^2,\qquad{\rm with}\quad n_\alpha\;\in\;\mathbb{N},
 \label{angularGen}\ee
but  the  bosonic generators $H_{\alpha}, S_{\alpha}, h_{\alpha\bar\beta}$, become  locally defined,
 $\varphi_\alpha\;\in \;[0, 2\pi/n_\alpha)$,  and fail to be constants of motion.
To get the globally defined bosonic generators we have to take their  relevant powers,
\be
{\widetilde H}_\alpha :=(H_{\alpha})^{n_\alpha},\quad
{\widetilde K}_{\alpha} := (K_{\alpha})^{n_\alpha},\quad
{\widetilde h}_{\alpha\bar\beta} :=(h_{\alpha\bar\beta})^{n_\alpha n_\beta}.
\ee
as well as replace the  fermionic generator $\Theta_{A\bar\alpha}$ by the  following one
\be
\widetilde{\Theta}_{A\bar\alpha} = (H_{\alpha})^{n_\alpha-1}\Theta_{A\bar\alpha}.
\ee
As a result, the dynamical (super)symmetry algebra becomes  nonlinear deformation of $su(1,N|M)$ .

The  angular Hamiltonian \eqref{angularGen}   defines  the  class
the superintegrable generalizations of the conformal mechanics, and of the oscillator- and Coulomb-like   systems on the $N$-dimensional Euclidean  spaces \cite{rapid}.
As a particular case, this class of systems includes the "charge-monopole" system \cite{monopole}, Smorodinsky-Winternitz system \cite{sw}(for the explicit expressions of the action-angle variables of these systems see, respectively,\cite{saghatel} and \cite{galajinsky}), as well as the rational Calogero models \footnote{To our best knowledge, action-angle variables for the angular part of the rational Calogero models are not yet constructed explicitly. However, we have at hand the spectrum of the  angular part of rational Calogero model \cite{flp}. Taking its  (semi)classical limit  we can conclude that it has the form \eqref{angularGen}, see, e.g.\cite{rapid}}. Thus, proposed systems can be considered as their $2M$ superconformal extensions.

 Since  the generators  $Q_A,S_A, R_{A\bar B}$ remain unchanged upon above identification (as well as the expression of the angular Hamiltonian \eqref{cas}   via  generators $H,K,D$), we conclude that listed generators  form   superconformal algebra  $su(1,1|N)$  with central charge \eqref{su11N}.\\

Finally, notice that  in \eqref{angularGen}  the nonzero constant $g\neq 0 $ appears,
and the range of validity of the action variables  is fixed to be  $I_\alpha\in [0,\infty)$.
As a result, standard free particle  and conformal mechanics cannot be included in the proposed description, since  for these systems we should choose $g=0, I_\alpha \in [0,\infty)$.
To exclude this constant we should replace the initial generators by the following ones
\be
\mathcal{H}:=H-\frac{g(g-2I )}{4K}, \qquad
\mathcal{H}_{\alpha}:={H}_{\alpha}+\imath g\frac{K_\alpha}{2K}, \qquad
\mathcal{Q}_{A}:= Q_{A}-\imath g\frac{S_{A}}{2K}.
\ee
This deformation will further ``non-linearize" the dynamical supersymmetry algebra $su(1,N|M)$.

\section{Oscillator- and Coulomb-like Systems}
In the previous section we mentioned that the angular Hamiltonian \eqref{angularGen} defines the superintegrable deformations of $N$-dimensional oscillator and Coulomb system \cite{rapid}, while in \cite{kns} the examples of such systems on  noncompact projective space $\widetilde{\mathbb{CP}}^N$ playing the role of phase space were constructed.
 So, one can expect that  on the
 phase superspace $\widetilde{\mathbb{CP}}^{N|M}$ one can construct the super-counterparts of that systems,
 which presumably, possess  (deformed) $\mathcal{N}=2M, d=1$ Poincar\'e supersymmetry.
Below we examine this question and show that our claim is corrects in some particular cases.

\subsection{Oscillator-like systems}
 We define the supersymmetric oscillator-like system by the
  the phase space   $\widetilde{\mathbb{CP}}^{N|M}$ (equipped with the Poisson brackets \eqref{pbb})
    by the   Hamiltonian
 \be
 H_{osc}=H+\omega^2K,
\label{osc} \ee
 where the generators $H,K$ are given by \eqref{b1}.
 In canonical coordinates  \eqref{cc}it reads
 \be
H_{osc} =\frac{p_r^2}{2}+\frac{(g+\sum_{\alpha=1}^{N-1}\pi_\a+ \sum_{A=1}^{M} \imath \bar\chi^A\chi^A )^2}{r^2}+\frac{\omega^2r^2}{2}.
 \ee
This system  possesses the $u(N)$ symmetry  given by the generators $h_{\alpha\bar\beta}$ defined in \eqref{b2}(among them $N-1$ constants of motion  $\pi_\alpha $ are
functionally independent), the $U(M)$ R-symmetry given by  the generators $R_{A\bar B}$ \eqref{b4} as well as  $N-1$ hidden symmetries given by the generators
 \be
M_{\alpha\beta}=(H_{\alpha}+\imath\omega K_\alpha )(H_{\beta}-\imath\omega K_\beta )=\frac{\bar{z}^\alpha \bar{z}^{\beta}}{A^2}(w^2+\omega^2)
\;:\quad \{H_{osc}, M_{\alpha\beta}\}=0,\label{M}\ee
The  generators \eqref{M} and the $su(N)$ generators $h_{\alpha\bar \beta}$  form the following symmetry algebra
\be
\{h_{\alpha\bar\beta}, M_{\gamma\delta}\}=\imath \left(M_{\alpha \delta}\delta_{\gamma\bar \beta}+M_{\gamma \alpha}\delta_{\delta\bar \beta} \right),
\quad
\{M_{\alpha\beta}, M_{\gamma\delta}\}=0,
\ee
\be
\{M_{\alpha\beta}, {\overline M}_{\gamma\delta}\}=\imath \left(4 \omega^2 I h_{\a\bar \delta}h_{\b \bar \g} -
\frac{M_{\a\b} \bar M_{\g \delta}} { h_{\a\bar \g}} \delta_{\a\bar \g}-
\frac{M_{\a\b} \bar M_{\g \delta}} { h_{\a\bar \delta}} \delta_{\a\bar \delta}-
\frac{M_{\a\b} \bar M_{\g \delta}} { h_{\b\bar \g}} \delta_{\b\bar \g}-
\frac{M_{\a\b} \bar M_{\g \delta}} { h_{\b\bar \delta}} \delta_{\b\bar \delta}
 \right),
\ee
with $I$ given by \eqref{casimir} and summation over repeated indices is not assumed.

Besides, this system has a  fermionic constants of motion  $\Theta_{A\bar \alpha}$  defined in \eqref{b3}.
Hence, it is superintegrable system in the sense of super-Liouville theorem, i.e. it has $2N-1$ bosonic and $2M$ fermionic, functionally independent, constants of motion \cite{shander}.
Further generalization to the systems with angular Hamiltonian \eqref{angularGen} is straightforward.\\

Let us show, that for the even $M=2k$ this system  possess the deformed  $\mathcal{N}=2k$ Poincar\'e supersymmetry, in the sense of papers  \cite{ivanovsidorov}.
For this purpose we choose the following Ansatz for supercharges
 \be
 \mathcal{Q}_A=Q_A+\omega C_{A B} {\bar S}_B,
 \label{tQ}\ee
 with    the constant matrix $C_{A B}$ obeying the conditions
 \be
 C_{A B}+C_{BA}=0,\qquad C_{AB}{\overline C}_{BD}=-\delta_{A\bar D}\label{C}
 \ee
 For sure, the condition  \eqref{C} assumes that $M$ is an even number, $M=2k$.

Calculating Poisson brackets of the functions \eqref{tQ} we get
\be
 \{\mathcal{Q}_A,\bar{\mathcal{Q}}_B\}=H_{osc} \delta_{AB},\qquad
 \{\mathcal{Q}_A,\mathcal{Q}_B\}=-\imath\omega\mathcal{G}_{AB}, \qquad
 \{ \bar{\mathcal{Q}}_A ,\bar{\mathcal{Q}}_B \}=\imath\omega\bar{\mathcal{G}}_{AB},
\ee
where
\be
 \mathcal{G}_{AB}:=
  C_{A C}R_{B\bar C} +C_{B C}R_{A\bar C} ,
\qquad \mathcal{G}_{\bar A\bar B}:= \bar{\mathcal{G}}_{AB}= \bar C_{AC}R_{C \bar B}+ \bar C_{BC} R_{C\bar A}, \qquad
\bar{\mathcal{G}}_{AB} = \bar C_{AC}\bar C_{DB}\mathcal{G}_{DC}.
\ee
Then we get that the algebra of generators $\mathcal{Q}_A$, $\mathcal{H}_{osc}$, $\mathcal{R}_A^B$ is closed indeed:
\begin{align}
&\{\mathcal{Q}_A,H_{osc}\}=\omega C_{AB} \mathcal{Q}_B, \qquad \{\mathcal{G}_{AB},H_{osc}\}=0, &\\
&\{\mathcal{Q}_A, \mathcal{G}_{BC}\}=
\imath(C_{AB}\mathcal{Q}_C+C_{AC}\mathcal{Q}_B), \qquad
\{\mathcal{Q}_A,\bar{\mathcal{G}}_{BC}\}=
-\imath(\bar C_{BD}\mathcal{Q}_D \delta_{A \bar C}+\bar C_{CD}\mathcal{Q}_D \delta_{A\bar B}).  &\end{align}

Hence, for the $M=2k$ the above oscillator-like  system \eqref{osc} possesses deformed $\mathcal{N}=4k$ supersymmetry. In the particular case $M=2$ the choice of the matrix $C_{AB}$ is unique(up to unessential phase factor):    $C_{AB}:={\rm e}^{\kappa}\varepsilon_{AB}$. In that case the  above relations   define the superalgebra $su(1|2)$-deformation of   $\mathcal{N}=4$ Poincar\'e supersymmetric mechanics studied in details  in  \cite{ivanovsidorov}.
For the $k\geq 2$  the choice of matrices $C_{AB}$ is not unique, and we get the  family of deformed $\mathcal{N}=4k$ Poincar\'e supersymmetric mechanics.
 \\

Let us present  other deformed $\mathcal{N}=2M$ Poincar\'e supersymmetric systems whose bosonic part is different from  those of  \eqref{osc}
 but nevertheless, has the oscillator potential.

For this purpose  we choose another Ansatz  for supercharges (in contrast with previous case $M$ is not restricted to be even number)
 \be
  \widetilde{\mathcal{Q}}_A=Q_A+\imath \omega S_A.
 \ee
These supercharges generates the  $su(1|M)$ superalgebra, and thus generalizes the systems considered in \cite{ivanovsidorov} to arbitrary $M$,
\begin{align}
&\{\widetilde{\mathcal{Q}}_A,\bar{\widetilde{{\mathcal{Q}}}}_B\}= \mathcal{H}_{osc}\delta_{AB}-\omega\mathcal{R}^A_B, \qquad\{\widetilde{\mathcal{Q}}_A,\widetilde{\mathcal{Q}}_B\}=0,\qquad \{\mathcal{R}_A^{\;B},\mathcal{R}_C^{\;D}\}=\imath (\mathcal{R}_A^{\;D}\de^B_C -\mathcal{R}_C^{\;B} \de^D_A)
&\label{oscQ1}\\
&
 \{ \widetilde{\mathcal{Q}}_A,\mathcal{R}^C_B\}=\imath\left(\frac{1}{M}\widetilde{\mathcal{Q}}_A \delta_{B \bar C}+
\widetilde{\mathcal{Q}}_B \delta_{A \bar C}\right) ,\qquad
\{ \widetilde{\mathcal{Q}}_A,\mathcal{H}_{osc}\}=\imath\omega\frac{2M-1}{M}\widetilde{\mathcal{Q}}_A,
&\label{oscQ2}
\end{align}
where
\be
\mathcal{H}_{osc}:= H_{osc} - \omega(I+\frac1M \sum_{C}R_{C\bar C}),\qquad \mathcal{R}_A^{\;B}:=R_{A\bar B}- \frac1M \delta_{A}^{B}\sum_{C}R_{C\bar C}
\label{tho}\ee
with  $I$ defined by \eqref{casimir}.
Hence, the Hamiltonian get the additional bosonic term proportional to the Casimir of conformal group.
In canonical coordinates \eqref{cc}  it reads
\be
\mathcal{H}_{osc}=\frac{p_r^2}{2}+\frac{\mathcal{I}}{r^2}+\frac{\omega^2 r^2}{2}
-\omega\left( \sqrt{2\mathcal{I}}+\frac{1}{M}(\bar \chi \chi)\right).
\ee
This Hamiltonian, seemingly, describes the oscillator-like systems specified by the  presence of external magnetic field.

So, choosing $\widetilde{\mathbb{CP}}^{N|M}$ as a phase superspace, we can  easily construct  superintegrable oscillator-like systems which possess deformed $\mathcal{N}=2M, d=1$ Poincar\'e supersymmetry.

\subsection{Coulomb-like systems}

Now, let us construct on the  phase space $\widetilde{\mathbb{CP}}^{N|M}$  with the Poisson bracket relations \eqref{pbb}, the Coulomb-like system   given   by the   Hamiltonian
 \be
 H_{Coul}=H+\frac{\gamma}{\sqrt{2 K}},
 \ee
 where the generators $H,K$ are defined by \eqref{b1}.

 The bosonic constants of motion of this system are given by the
 $u(N-1)$ symmetry generators  $h_{\a\b}$ , and by the $N-1$ additional constants of motion
\be
   R_{\alpha}=H_{\alpha}+\imath \gamma\frac{ K_{\alpha }}{I\sqrt{2 K}}\;:\quad\{H_{Coul},R_{\alpha}\}=\{H_{Coul},h_{\alpha\bar \beta}\}=0,
\label{Coul}\ee
where $H_\alpha, K_\alpha,\h_{\alpha\bar\beta}$ are defined  by \eqref{b2}.
These generators form the algebra
\be
\{R_\alpha,\bar R_{\bar\beta}\}=-\imath\delta_{\alpha \bar \beta}\Bigg(H_{Coul}-\frac{\imath\gamma^2}{2I^2}\Bigg)+\frac{\imath\gamma^2 h_{\alpha\bar\beta}}{2I^3},\quad
\{h_{\alpha\bar\beta},R_\gamma\}=\imath \delta_{\gamma \bar \beta}R_{\alpha},\quad
\{R_\alpha,R_\beta\}=0.
\ee
Besides, proposed system   has  $2M$ fermionic constants of motion given by $\Theta_{A\bar\alpha}$, and  $u(M)$ R-symmetry given by $R_{A\bar B}$.
Hence, it   is superintegrable in the sense of super-Liouville theorem \cite{shander}.
So, we constructed the maximally superintegrable Coulomb problem with dynamical $SU(1,N|M)$ superconformal symmetry which inherits all symmetries of initial bosonic system. \\

One can expect, that in analogy with oscillator-like system, our Coulomb-like system would possess (deformed) $\mathcal{N}=2M$-super-Poincar\'e symmetry
for $M=2k$ and $\gamma >1$. However, it is not a case.

Indeed, let us  choose the following Ansatz for supercharges
 \be
 \mathcal{Q}_A=Q_A+\sqrt{2\gamma} C_{A B}\frac{ {\bar S}_B}{(2K)^{3/4}},
 \ee
 with    the constant matrix $C_{A B}$ obeying the conditions \eqref{C}, $M=2k$ and $\gamma>0$.

 Calculating their Poisson brackets we find
\begin{align}
 &\{\mathcal{Q}_A,\bar {\mathcal{Q}}_B\}= H_{Coul} \de_{A\bar B}+\frac{3}{2}\frac{\sqrt{2\g}}{(2K)^{7/4}}\left(
 S_A\bar C_{BD} S_D +\bar S_B C_{AD}\bar S_D
 \right),&\\
 &
 \{\mathcal{Q}_A,\mathcal{Q}_B\}=
 -\frac{\imath\sqrt{2\g}}{2(2K)^{3/4}}(C_{BD}\mathcal{R}_{A}^D+C_{AC}\mathcal{R}_{B}^D),\quad \{\mathcal{Q}_A, \mathcal{R}_{B}^{\;C}\}=-\imath \mathcal{Q}_{B}\de_{A\bar C} ,
\end{align}
where $\mathcal{R}^A_B$ is defined in \eqref{tho}.

Further calculating the Poisson brackets of $\mathcal{Q}_A$ with the generators appearing in the r.h.s. of the above expressions we get that the superalgebra  is not closed. For example,
\be
\{\mathcal{Q}_A, H_{Coul}\}=\frac{3\g}{(2K)^{3/2}} S_A+\frac{\sqrt{2\g}}{(2K)^{3/4}}C_{AB}\left( \mathcal{\bar Q}_B -\frac{3}{4K}
 \bar S_B D \right).
\ee
Hence, proposed supercharges do not yield closed deformation of   $\mathcal{N}=2M$-super-Poincar\'e algebra.\\

Let us choose another Ansatz for supercharges (as above we assume that $\gamma >0$)
 \be
  \widetilde{\mathcal{Q}}_A=Q_A+ \imath\sqrt{2\gamma} e^{\imath \frac{\pi}{2}}\frac{S_A}{(2K)^{3/4}},
 \ee
which yields
\be
\{\widetilde{\mathcal{Q}}_A,\bar{\widetilde{\mathcal{Q}}}_B\}= \mathcal{H}_{Coul}\de_{A\bar B}+
\frac{\sqrt{2\g}}{2(2K)^{3/4}}\mathcal{R}^B_A , \qquad
\{\widetilde{\mathcal{Q}}_A,\widetilde{\mathcal{Q}}_B\}=0,\qquad
\{\widetilde{\mathcal{Q}}_A,\mathcal{R}^C_B \}
=\imath\left(\frac1M \widetilde{\mathcal{Q}}_A\de_{B \bar C} -\widetilde{\mathcal{Q}}_B\de_{A\bar C}\right),
\ee
where
\be
\mathcal{H}_{Coul}=H_{Coul}-\frac{\sqrt{2\g}}{(2K)^{3/4}}\left( I-\frac{1}{2M}\sum_C R_{C\bar C}\right),
\ee
with  $I$ and $\mathcal{R}^A_B$ are defined, respectively,  in \eqref{casimir2} and \eqref{tho}.
In canonical coordinates \eqref{cc} this Hamiltonian reads
\be
\mathcal{H}_{Coul}=\frac{p_r}{2}+\frac{\mathcal{I}}{r^2}+\frac{\gamma}{r}-\frac{\sqrt{2\gamma}}{r^{3/2}}
\left(g+\sum_\alpha \pi_\alpha+\frac{2M-1}{2M}(\bar \chi \chi)\right).
\ee
However, one can easily check that proposed supercharges do not yield closed deformation of Poincar\'e superalgebra as well, e.g.
\be
\{ \widetilde{\mathcal{Q}}_A, \frac{ \mathcal{R}^C_B }{(2K)^{3/4}}\}=
\frac{\imath}{ (2K)^{3/4}}\left(\frac1M \widetilde{\mathcal{Q}}_A\de_{B \bar C} -\widetilde{\mathcal{Q}}_B\de_{A\bar C}\right)+\frac{3 }{2}\frac{S_A}{(2K)^{7/4}}\mathcal{R}_B^C
\ee

So, proposed superextensions of Coulomb-like systems, being well-defined from the viewpoint of superintegrability, do not possess neither $\mathcal{N}=2M$ supersymmetry, no its deformation.
The $su(1,N|M)$ superalgebra plays the role of dynamical algebra of that systems.

\section{  Fubini-Study-like K\"ahler structure}
The above considered super-Ka\"hler structure is  obviously the higher-dimensional super-analogue of the Klein model of Lobachevsky space.
On the other hand, Lobachevsky space has other common parametrization as well, which is known as Poincar\'e disc \cite{dnf}.
The  higher-dimensional generalization of Poincar\'e disc parameterizing the noncompact complex projective space is quite similar to the Fubini-Study structure for $\mathbb{CP}^N$. It is defined by the K\"ahler potential
\be
\mathcal{K}=-g\log( 1-\sum_{a=1}^N z^a\bar z^a).
\ee
For the obtaining of the super-analogue of this potential from $\mathbb{C}^{1,N|M}$, one should pass from the matrix \eqref{mat0} to the diagonal
  matrix
$\gamma_{a\bar b}=\diag{1,-1,\ldots,-1}$. This   can be done  by the transformation
\be
v^0\to\frac{v^0+v^N}{\sqrt{2}},\qquad v^N\to \frac{v^0-v^N}{\imath \sqrt{2}}.
\label{K-P}
\ee

On the  reduced phase space \eqref{K-P} corresponds
to the transformation
\be
w\to\imath\frac{z^N-1}{z^N+1}, \qquad z^\a \to \sqrt{2}\frac{z^\a}{z^N+1},\qquad\th^A\to\sqrt{2}\frac{\th^A}{z^N+1}.
\ee
Thus  we will get the Fubini-Study-like K\"ahler potential
\be
\mathcal{K}=-g \log(1-z^c \bar z^c + \imath \th^C \bar \th^C),
\label{PoinPot}\ee
which defines the following K\"ahler structure
\be
\Omega= \frac{\imath}{g}\left[\left(\frac{g \de_{a\bar b}}{\tilde{A}}+
\frac{\bar z^a z^b}{\tilde{A}^2}\right) dz^a \wedge d\bar z^b +
\frac{\imath\bar \th^A z^a}{\tilde{A}^2} d\th^A \wedge d\bar z^a
-\frac{\imath \bar z^a \th^A}{\tilde{A}^2} dz^a\wedge d\bar \th^A -
\left(\frac{g\de_{A\bar B}}{\tilde{A}}+\frac{\bar \th^A \th^B}{\tilde{A}^2}\right)d\th^A \wedge d\bar \th^B
\right],
\ee
%
where we have used a similar notation as in \eqref{A}
\be
\tilde{A}:=\frac{1-z^c \bar z^c + \imath \th^C \bar \th^C}{g}.
\label{Atilde}\ee
The respective Poisson brackets read:
\be
\{z^a,\bar z^b\} =\imath {\tilde A}\left( {\de^{a\bar b}-z^a\bar z^b  }\right), \qquad \{z^a,\bar \th^A\}=\imath {\tilde A} {z^a \bar \th^A}, \qquad
\{\th^A, \bar \th^B\}= {\tilde A}\left({\de^{A\bar B}+\th^A\bar \th^B  }\right).
\ee

Now let us introduce the  canonical coordinates, but now taking the symplectic/K\"ahler one form associated with the K\"ahler potential \eqref{PoinPot}, i.e. the one that define "Fubini-Study"-like metric. Then, as before, one needs to identify it with the canonical one, and this canonical coordinates will play the role of "Cartesian" coordinates instead of the "spherical" ones discussed above.
\be
\tilde{\mathcal{A}}= -\frac{g}{2}\frac{\imath (\bar z^a d{z}^a -{z}^a d \bar {z}^a)+ {\th}^A d\bar {\th}^A +\bar {\th}^A d{\th}^A}{1-
{z}^c \bar {z}^c +\imath {\th}^C \bar {\th}^C}
:= p_ad\varphi_a +\frac{1}{2}{\chi}^A d\bar {\chi}^A + \frac{1}{2}\bar {\chi}^A d{\chi}^A.
\ee
It leads to the relations
\be
{z}^a = \sqrt{\frac{p_a}{g+p-\imath {\chi}^C \bar {\chi}^C}} e^{\imath \varphi_a}, \qquad
{\th}^A = \frac{\sqrt{2}}{r}{\chi}^A, \qquad p=\sum_a p_a \;,
\label{fsc}\ee
 or
\be
p_a=\frac{z^a\bar z^a}{\tilde{A}},\qquad  \varphi_a=\arg (z^a),\qquad \chi^A=\frac{\th^A}{\sqrt{\tilde{A}}},
\ee
where $\tilde{A}$ is defined by \eqref{Atilde}.

These coordinates are related with \eqref{cc}   as follows:
\be
p_\a =\pi_\a,\quad p_N= \frac{1}{4}\left(p^2_r + \left(r-\frac{\sqrt{2\mathcal{I}}}{r}\right)^2\right), \quad
\varphi_N = \arctan\left(\frac{2xy}{(x-y)(x+y)}\right),
\ee
where
\be
x=1-\frac{p_r^2}{r^2}-\frac{2\mathcal{I}}{r^4}, \quad y=\frac{p_r}{r},
\ee
while $  {\chi}^A$ and $\varphi_\alpha$ remains unchanged after transition from one parameterization  to the other.

Finally, let us draw readers  attention to the complete similarity of the bosonic part of  \eqref{fsc} with the equations mapping parameterizing
compactified Ruijsenaars-Schneider model with excluded centre of mass to the complex projective (phase) space $\mathbb{CP}^N$. This prompt us, at first, to construct the  conformal-invariant analogue of that model by replacing the complex projective space by its noncompact analogue $\widetilde{\mathbb{CP}}^N$. Then one can try to construct its $su(1,N|M)$-supeconformal extension by further replacement of $\widetilde{\mathbb{CP}}^N$ by $\widetilde{\mathbb{CP}}^{N|M}$.

\section{Concluding remarks}
In this paper we  suggested to construct  the $su(1,N|M)$-superconformal mechanics
 formulating them on  phase superspace given by the non-compact analogue of complex projective superspace $\widetilde{\mathbb{CP}}^{N|M}$.  The $su(1,N|M)$ symmetry generators   were defined there  as  a Killing potentials of $\widetilde{\mathbb{CP}}^{N|M}$. We parameterized this phase space   by the specific coordinates allowing to interpret it as a higher-dimensional super-analogue of the Lobachevsky plane parameterized by lower half-plane (Klein model). Then we transited to the canonical coordinates  corresponding to the known separation of  the "radial" and "angular" parts of (super)conformal mechanics. Relating the "angular" coordinates with action-angle variables we demonstrated that proposed scheme allows to construct the $su(1,N|M)$ supeconformal extensions of wide class of superintegrable systems.  We also proposed  the superintegrable oscillator- and Coulomb- like systems   with a $su(1,N|M)$ dynamical superalgebra, and found that oscillator-like systems admit  deformed $\mathcal{N}=2M$ Poincar\'e supersymmetry, in contrast with Coulomb-like ones.

 In fact, proposed scheme  demonstrated the effectiveness  of the supersymmetrization via  formulation of the initial systems in terms of K\"ahler phase space and   further superextension of the latters.
 In order to relate considered systems with the conventional ones (with Euclidean configuration spaces), we restricted ourself by the non-compact complex projective superspace. So, we are sure that applying the same approach to the conventional (compact) complex projective spaces we can find many new integrable systems as well  and construct their unexpected  extended supersymmetric extensions.

 Proposed scheme could  obviously be extended to the systems on complex Grassmanians   (and on their noncompact analogues). In particular, we expect to find, in this way, the $\mathcal{N}$-supersymmetric extensions of compactified spin-Ruijsenaars-Schneider models.
 Moreover,  it seems to be straightforward task to  apply proposed approach to the systems with generic $U(N)$-invariant K\"ahler phase spaces locally defined by the K\"ahler potential $\mathcal{K}\left( z^a\bar z^a\right)$. We expect  that it can be done in terms of K\"ahler phase superspace locally defined by the potential
 \be
 \widetilde{\mathcal{K}}=\mathcal{K}\left( 
 z^a\bar z^a+\imath
 \eta^a\bar\eta^A\right).
 \ee
 In this way we expect to construct the $\mathcal{N}=2M$ supersymmetric  extensions of the systems with curved (Riemann) configuration space as well, in particular, of the so-called $\kappa$-deformations (i.e. spherical/hyperbolic generalizations) of conformal mechanics, oscillator and Coulomb systems\cite{ranada,shmavon}.

 Finally, notice that considered phase superspace is not associated with external algebra of initial bosonic manifold, and thus, it is not related  with the superfield approach.
 Thus,  it is interesting to consider the systems with $(N|kM)$-dimensional  K\"ahler phase superspaces defined by the potentials
 \be
  \widetilde{\mathcal{K}}=\mathcal{K} \left( 
   z^a\bar z^a\right)+ F \left(\imath
   g_{a\bar b}\eta^a_\alpha\bar\eta^b_\alpha \right) ,\qquad F'(0)={\rm const},
 \ee
 and construct, in this way, the $\mathcal{N}=kN$ supersymmetric mechanics.
 Very preliminary attempt  in this direction was done in \cite{npps} where the $\mathcal{N}=2$ supersymmetric extensions of the systems with generic K\"ahler phase space was considered. However, this promising direction was not further developed since that time.
 We plan to consider listed problems elsewhere.

\acknowledgements

This work was supported by  the Russian Foundation
of Basic Research   grant 20-52-12003 (S.K., A.N.) and
by the Armenian Science Committee  projects   20RF-023, 21AG-1C062 (E.Kh., A.N.)   and 21AA-1C001(E.Kh.).
 The work  of E.Kh. was completed  within the Regional Doctoral Program on Theoretical and Experimental Particle Physics Program sponsored by VolkswagenStiftung, and within ICTP Network Program NT-04.

\end{document}